\documentclass[showpacs,preprintnumbers,superscriptaddress,amsmath,amssymb,floatfix,aps]{revtex4}

\usepackage[dvips]{graphicx}
\usepackage{dcolumn}
\usepackage{bm}
\usepackage{ulem}

\begin{document}
\title{Radiation induced force between two planar waveguides}

\author{Francesco Riboli}
\affiliation{LENS, via N. Carrara 1, I-50019 Sesto Fiorentino, Firenze, Italy}
\author{Alessio Recati}
\affiliation{CNR-INFM Center on Bose-Einstein Condensation and Dipartimento di Fisica, Universit\'a di Trento, via Sommarive 14, I-38050 Povo, Trento, Italy}
\author{Mauro Antezza}
\affiliation{CNR-INFM Center on Bose-Einstein Condensation and Dipartimento di Fisica, Universit\'a di Trento, via Sommarive 14, I-38050 Povo, Trento, Italy}
\author{Iacopo Carusotto}
\affiliation{CNR-INFM Center on Bose-Einstein Condensation and Dipartimento di Fisica, Universit\'a di Trento, via Sommarive 14, I-38050 Povo, Trento, Italy}

\begin{abstract}
We study the electromagnetic force exerted on a pair of parallel slab
waveguides by the light propagating through them.
We have calculated the dependence of the force on the slab separation by means of the Maxwell--Stress tensor formalism and we have discussed its main features for the different propagation modes:  spatially symmetric (antisymmetric) modes give rise to an attractive (repulsive) interaction. 
We have derived the asymptotic behaviors of the force at small and large separation and we have quantitatively estimated the mechanical deflection induced on a realistic air-bridge structure.
\end{abstract}


\maketitle

\section{Introduction}

When two objects in close proximity are illuminated by a light source an optical force is exerted on each of them, 
whose sign can be either attractive or repulsive depending on the geometry of the objects and of the optical
mode in which light propagates.
Physically, this force originates from the interaction of the induced
dipoles in the dielectric media by the electromagnetic field of the light wave.
Its magnitude is proportional to the light intensity and depends on the actual 
profile of the electromagnetic field and on its polarization.

The recent advances of nanotechnologies have led to the realisation
of solid-state samples whose sizes and separations are so small that
the optical forces can have a significant impact on the shape and the 
position of the object.
In particular, one may expect in the next future to take advantage of 
these optical forces to control the growth and the self-assembling of artificial 
materials such as photonic crystals or random sphere assembly.
Moreover optical forces may be also used to engineer 
the quantum state of the mechanical motion of nano-objects~\cite{wilson-rae}.

In the last years, a significant amount of experimental and theoretical studies
have concerned coupled resonant systems, such as spherical and disk-shaped whispering gallery resonators \cite{resonator} as well as double-layer photonic-crystal-slab 
cavities \cite{Notomi}. 
Thanks to the extremely high quality factor of these systems, the
light intensity in the resonator is in fact strongly enhanced as
compared to the input power, which leads to a corresponding increase
in the magnitude of the force. 

At the same time the case of two coupled air bridge silicon waveguides
with square cross section has been investigated in~\cite{povinelli}:
from the reported numerical calculations, it turns out that the displacement
of the silicon wire-bridges under the action of the induced
force can reach values measurable with standard Atomic Force
Microscope techniques already at reasonable input laser power.

Motivated by this intense research and by the fast advances in the
nanotechnological expertise in manipulating these systems, we here
present a systematic characterization of the optical force between two  
parallel planar waveguides. 
For such a simple geometry most of the results can be obtained by analytical means, which 
provides useful insight into the basic physics of the optical forces. 

The paper is organized as follows. In Sec.~\ref{SecII} the
Maxwell equations for two coupled waveguides system are solved, and, in particular, 
expressions are obtained for  the fields and the dispersion relations.
In Sec. \ref{sec:force} the Maxwell stress tensor technique is used to calculate the optical force,  whose behavior is studied in detail as a function of the
distance between the waveguides and of the incident light frequency. In Sec. \ref{airbridge} we quantitatively estimate the mechanical deflection produced by the radiation force in a realistic air-bridge device and we show that it is strong enough to be experimentally measured by means of Atomic Force Microscopy techniques.   
Finally, in Sec. \ref{sec:limits} we derive closed expressions for the force in the limiting cases of small and large separation.

\section{Guided electromagnetic modes in coupled slab waveguides} 
\label{SecII}

\begin{figure}[ptb]
\begin{center}
\includegraphics[width=0.7\textwidth]{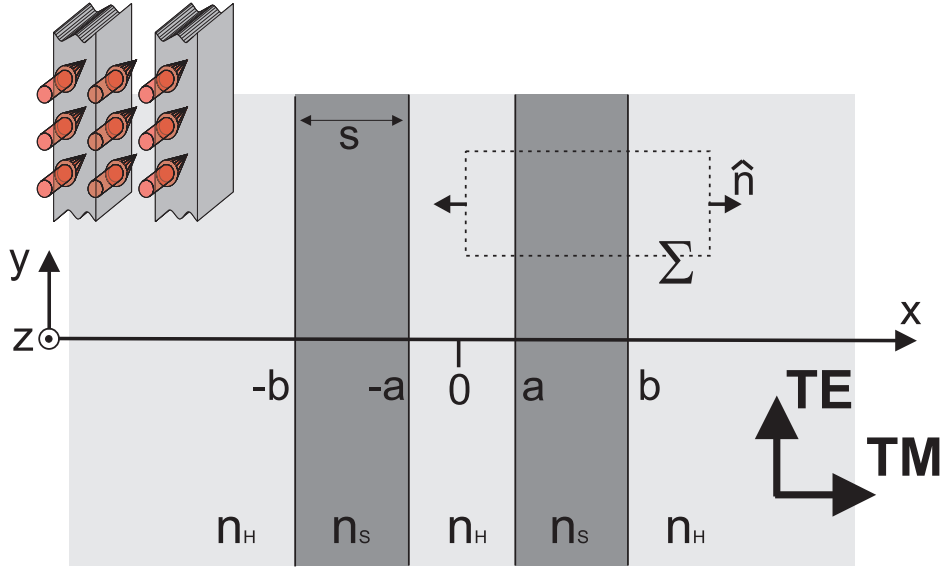}
\caption{\footnotesize Left panel: Schematic view of the two coupled waveguide system
 under study. The red  arrows represent the input laser light. 
Right panel: Transverse cut of the system along a fixed-$z$ plane. 
The $TE/TM$ arrows indicate the polarization of the electric field in
 the two polarization states.}
\label{Schematic_problem}
\end{center}
\end{figure}

The physical configuration we consider in the present paper is shown
in Fig.\ref{Schematic_problem}: a pair of parallel dielectric slabs of
thickness $s$ and refractive index $n_S$ separated by a distance $2a$
and embedded in a host medium of lower refractive index $n_H<n_S$.
The axis $x$ is orthogonal to the slabs and light is assumed (without
loss of generality) to be in a plane wave state propagating along $z$.
Under this assumption, the electric and magnetic fields can be written
in the form: 
\begin{eqnarray}
{\bf E}({\bf r},t)=\textrm{Re}\left[{\bf E}(x)\,e^{i(\omega
t-\beta z)} \right], \label{Harmonic_solutions1}\\
{\bf H}({\bf r},t)=\textrm{Re}\left[{\bf H}(x)\,e^{i(\omega
t-\beta z)}\right], \label{Harmonic_solutions2}
\end{eqnarray}
$\beta$ being the wave number of the propagation along $z$ and $\omega$
the angular frequency. 
The $y$ dependence of the fields has disappeared thanks to
the translational invariance of the system in the waveguide plane, and
to the choice made for the direction of propagation.
The general solution for the electromagnetic fields can be obtained as a
linear combination of such plane waves.

Depending on the orientation of the fields with respect to the
propagation direction, two independent electromagnetic
polarizations states can be identified, known as Transverse
Electric ($TE$) and Transverse Magnetic ($TM$)~\cite{Marcuse,Okamoto}. 
The $TE$ polarization state is characterized by $E_x=E_z=H_y=0$, 
while the $TM$ polarization is characterized by $E_y=H_x=H_z=0$

For each $TE/TM$ polarization state, the field wavefunction is
determined by solving the corresponding Maxwell equations.
Since we are considering modes which are guided by the waveguide system,
the electromagnetic field is confined in the slabs, where the wave-vector along the $x$-direction is purely real $k_x=\pm\kappa=\pm\sqrt{k^2n^2_S-\beta^2}$, with $k=\omega/c$. In the surrounding host medium the electromagnetic field is evanescent with a purely imaginary wave-vector $k_x=\pm i\sigma=\pm i\sqrt{\beta^2-k^2n^2_H}$.

For the $TE/TM$ polarizations, this is summarized in a field
wavefunction (that is the $E_y$ or $H_y$ fields for the $TE/TM$ modes,
respectively) which reads:
\begin{equation}
\!\!\!\!\!\!\!\!\!\!\!\!\!\!\!\!\!\!\!\!\!\!\!\!\!\!\!\!\!\!\!\!\!\!\!\!E_{y},H_{y}=A\,
\left\{%
\begin{array}{ll}
    \alpha_1\,e^{-\sigma(x-a-s)} \hspace{4cm}          & \hbox{$(x>a+s)$} \\
    \cos [\kappa(x-a)+\phi_2]              & \hbox{$a\leq x\leq a+s$} \\
    \alpha_2\,\cosh(\sigma x)+\alpha'_2\,\sinh(\sigma x)    & \hbox{$-a\leq x\leq a$} \\
    \cos [\kappa(x+a)-\phi_4]              & \hbox{$-a\leq x\leq -a-s$} \\
    \alpha_3\,e^{+\sigma(x+a+s)}                       & \hbox{$x<-a-s$}, \\
\end{array}%
\right. .     \label{5slab_soluction1}
\end{equation}
The system having reflection symmetry with respect to the $x=0$ plane,
the electromagnetic modes can be classified as symmetric and
antisymmetric depending on the symmetry operation of the field
wavefunction, namely $E_y$ for the $TE$ modes or $H_y$ for the $TM$
ones.
Note that this classification is related, but not identical to the
usual one in terms of the reflection symmetry of the full
electromagnetic field, where the magnetic ${\bf H}$ field transforms
as a pseudovector, while ${\bf E}$ is instead a vector.

The dispersion relation connecting $\beta$ to $\omega$, as well as a
relation between the amplitude coefficients $\alpha$ and the phases
$\phi$ in the different regions are then obtained by matching the field (\ref{5slab_soluction1}) in the different regions according to the symmetry of the electric and magnetic fields.
For the symmetric modes, the dispersion law has the following
analytical, yet implicit form:
\begin{eqnarray}
\kappa s= \arctan \left(\frac{\sigma}{\kappa}\right)+\arctan
\left[\frac{\sigma}{\kappa}\tanh(\sigma a)\right]-m\pi,  
 &&\;\;\;\;\textrm{{\sl TE}} \label{TE_modes_symm}\\
\kappa s= \arctan
\left(\frac{n^2_S}{n^2_H}\frac{\sigma}{\kappa}\right)+\arctan
\left[\frac{n^2_S}{n^2_H}\frac{\sigma}{\kappa}\tanh(\sigma
a)\right]-m\pi, 
&&\;\;\;\; \textrm{{\sl TM}}
\label{TM_modes_symm}
\end{eqnarray}
for the $TE/TM$ polarizations, respectively.
The dispersion laws for the anti-symmetric modes are
obtained from the symmetric ones  by replacing
$\tanh(x)\mapsto 1/\tanh(x)$.
In the following, we shall see that this fact holds for other physical
quantities as well. In all these dispersion laws, the (integer) quantum number 
$m\geq 0$ specifies the number of nodes of the wavefunction inside
each slab.

In summary, the optical modes propagating along $z$ in a given two
waveguide system are classified by their $TE/TM$ polarization state,
the symmetric/anti-symmetric character of the field wavefunction with
respect to reflections on the $x=0$ plane, and the number $m\geq 0$ of
nodes inside each waveguide.
Thanks to the scaling properties of the Maxwell equations, the
qualitative shape of the dispersion relations
depends on the geometrical
parameter $a/s$ only, which quantifies the ratio between the spacing $a$
and the thickness $s$ of each slab.
The absolute value of $s$ fixes the natural frequency scale
$\omega_s=2\pi\,c/s$.

\begin{figure}[ptb]
\begin{center}
\includegraphics[width=1\textwidth]{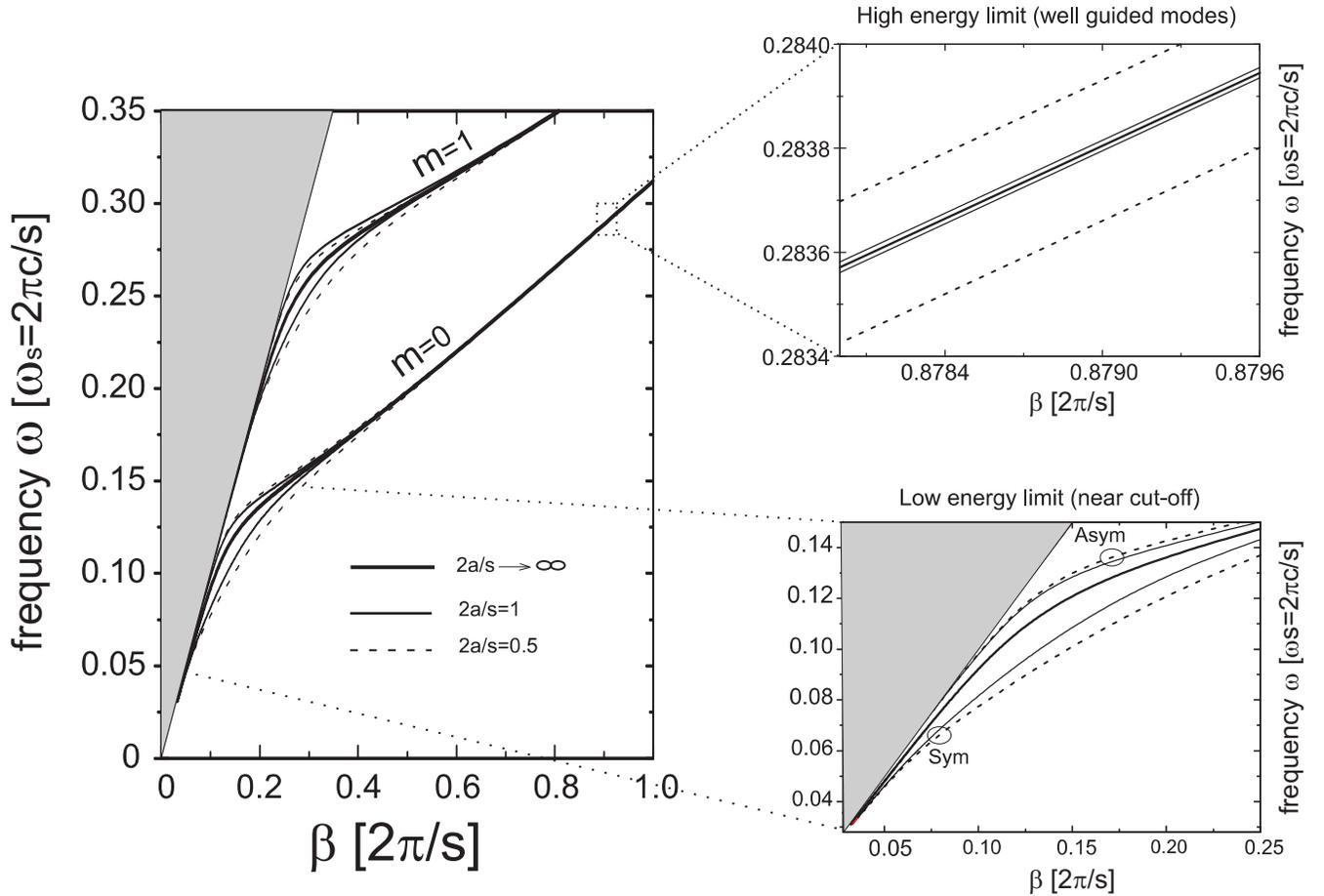}
\caption{\footnotesize Main panel: dispersion relation of the different symmetric and
 antisymmetric branches of the fundamental ($m=0$) and first
 order ($m=1$) $TM$ mode for various waveguides separations. 
 The thick continuous lines refer to the infinite separation 
$a/s\rightarrow \infty$ case where symmetric and anti-symmetric modes
 are degenerate.
Upper and lower lines with respect to the infinite separation one refer to symmetric (Sym) and
 antisymmetric (Asym) modes respectively; the thin continuous lines are for $a/s=0.5$, while the dashed ones are for a shorter separation $a/s=0.25$.
In the smaller panels on the right we have highlighted the behavior of
 the $m=0$ modes, respectively far from the cut-off (high energy
 limit) and near the cut-off (low energy limit).}
\label{Band_diagram1}
\end{center}
\end{figure}

In Fig.~\ref{Band_diagram1}, we have plotted the dispersion of the
different modes ($TE/TM$, symmetric and anti-symmetric) of the
coupled slab waveguide system for different values of the
geometrical parameter $a/s$.
All the branches have a lower cut-off to the frequencies that can actually propagate in a given mode of the coupled waveguide system.
At the cut-off point, the waveguide dispersion coincides with the free photon
dispersion $\omega=c\beta/n_H$ of the host medium.

At infinite separation $a/s=\infty$, the dispersion reduces to the one
of an isolated waveguide, and for every polarization state the
symmetric and anti-symmetric branches are degenerate. 
As the two waveguides are pushed closer, this degeneracy is lifted, and
every branch experiences a frequency shift whose sign depends on
its symmetric or anti-symmetric nature.
As usual for the bonding/anti-bonding electronic states in diatomic
molecules~\cite{CCT_MQ}, the symmetric states are pushed toward
lower frequencies by the coupling, while the anti-symmetric ones are
pushed toward higher frequencies.
As a consequence, the cut-off frequency experiences itself a
shift of the same sign and comparable magnitude.

In the next sections, we shall study the electromagnetic pressure acting
on each of the two slab waveguides because of the presence of the 
other slab.
The pressure being proportional to the intensity of light propagating
along the waveguide system, it is important to relate the amplitude
coefficient $A$ in (\ref{5slab_soluction1}) to the power density $P$
for unit length in the $y$-direction. 
This is easily calculated from the flux of the Poynting vector through a
planar section orthogonal to the propagation direction. 

For the symmetric modes, we obtain
\begin{eqnarray}
\!\!\!\!\!\!\!\!\!\!\!\!\!\!\!\!\!\!\!\!\!\!\!\!\!\!\!\!\!\!\!\!\!P=\frac{|A_{TE}|^2\,\beta
s}{2\omega\mu_0}\left[\frac{1-\tanh^2(\sigma
a)}{1+\frac{\sigma^2}{\kappa^2}\tanh^2(\sigma
a)}\frac{a}{s}+1+\frac{1}{\sigma
s}\left(1+\frac{(1+\frac{\sigma^2}{\kappa^2})\tanh(\sigma
a)}{1+\frac{\sigma^2}{\kappa^2}\tanh^2(\sigma a)}\right)
\right],\label{P_TE} \\
\!\!\!\!\!\!\!\!\!\!\!\!\!\!\!\!\!\!\!\!\!\!\!\!\!\!\!\!\!\!\!\!\!P=\frac{|A_{TM}^2|\,\beta
s}{2\omega\epsilon_0n^2_{_S}}
\left[\frac{1-\tanh^2(\sigma
a)}{1+\frac{n^4_{_S}}{n^4_{_H}}\frac{\sigma^2}{\kappa^2}\tanh^2(\sigma 
a)}\frac{a}{s\frac{n^2_{_S}}{n^2_{_H}}}+1+\frac{1}{\sigma
s\frac{n^2_{_H}}{n^2_{_S}}}\left(\frac{1+
\frac{\sigma^2}{\kappa^2}}{1+\frac{n^4_{_S}}{n^4_{_H}}
\frac{\sigma^2}{\kappa^2}}+\frac{(1+\frac{\sigma^2}{\kappa^2})\tanh(\sigma
a)}{1+\frac{n^4_{_S}}{n^4_{_H}}\frac{\sigma^2}{\kappa^2}\tanh^2(\sigma
a)}\right) \right].\label{P_TM}
\end{eqnarray}
for the $TE/TM$ polarizations, respectively. The expressions for the corresponding antisymmetric modes are obtained by replacing $\tanh(x)\mapsto
\tanh(x)^{-1}$ in (\ref{P_TE}) and (\ref{P_TM}).

\section{Force between two planar waveguides}
\label{sec:force}

Starting from the electromagnetic field profiles discussed in the
previous section, we shall now proceed with a calculation of the average
electromagnetic force ${\bf F}$ acting on the slabs when a monochromatic
wave of frequency $\omega$ is propagating along the coupled waveguide
system in a well-defined mode.
To keep the treatment as simple as possible, we shall assume this force
to be equilibrated by some other, unspecified, force which keeps the
system at mechanical equilibrium at all times. 
The calculation of the force will then be performed in the framework of
the macroscopic electrodynamics of continuous media using the Maxwell
stress tensor $T$~\cite{Jackson,LLFT}. 
This technique allows one to directly calculate the force, and has been
extensively used in the literature to estimate forces of electromagnetic
origin in many contexts, from clusters of dielectric spheres~\cite{Pendry}
to waveguides~\cite{povinelli}, to quantum fluctuations as in the Casimir
effect~\cite{LLCondMat}. 
An important point of this formalism is that it does not make any
assumption on the microscopic nature of the material media under
examination and therefore can easily take into account absorption
effects.
In the following we shall focus our attention on the case of
dielectric slabs with a real refraction index $n_S=n$ embedded in air,
for which $n_H\simeq 1$. 

As all the fields have a monochromatic time dependence at frequency
$\omega$, momentum conservation arguments show that the average
electromagnetic force acting on a body is equal to the surface
integral of the time averaged Maxwell tensor  
\begin{equation}
\bar{T}_{ij}=\frac{\epsilon_0}{2}{\rm Re}
\left[
E_i E_j^*+ H_i H_j^*-\frac{1}{2}\,\delta_{ij}\,\sum_k \left(E_k E_k^*
+ B_k B_k^* \right)
\right], 
\end{equation}
over an arbitrarily chosen closed surface $\Sigma$ enclosing the body
with outward orientation:
\begin{equation}
{\bf F}_i=\int_{\Sigma} \bar{T}_{ij}(\mathbf{r})
\,\textrm{d}\mathbf{\sigma}_j, \label{Fmst}
\end{equation}
In our specific configuration, a good choice for $\Sigma$ is the one
 shown in Fig.\ref{Schematic_problem}b, that is a cylinder of axis
 parallel to $z$ and with rectangular bases parallel to the $xy$ plane.

\begin{figure}[ptb]
\begin{center}
\includegraphics[width=1\textwidth]{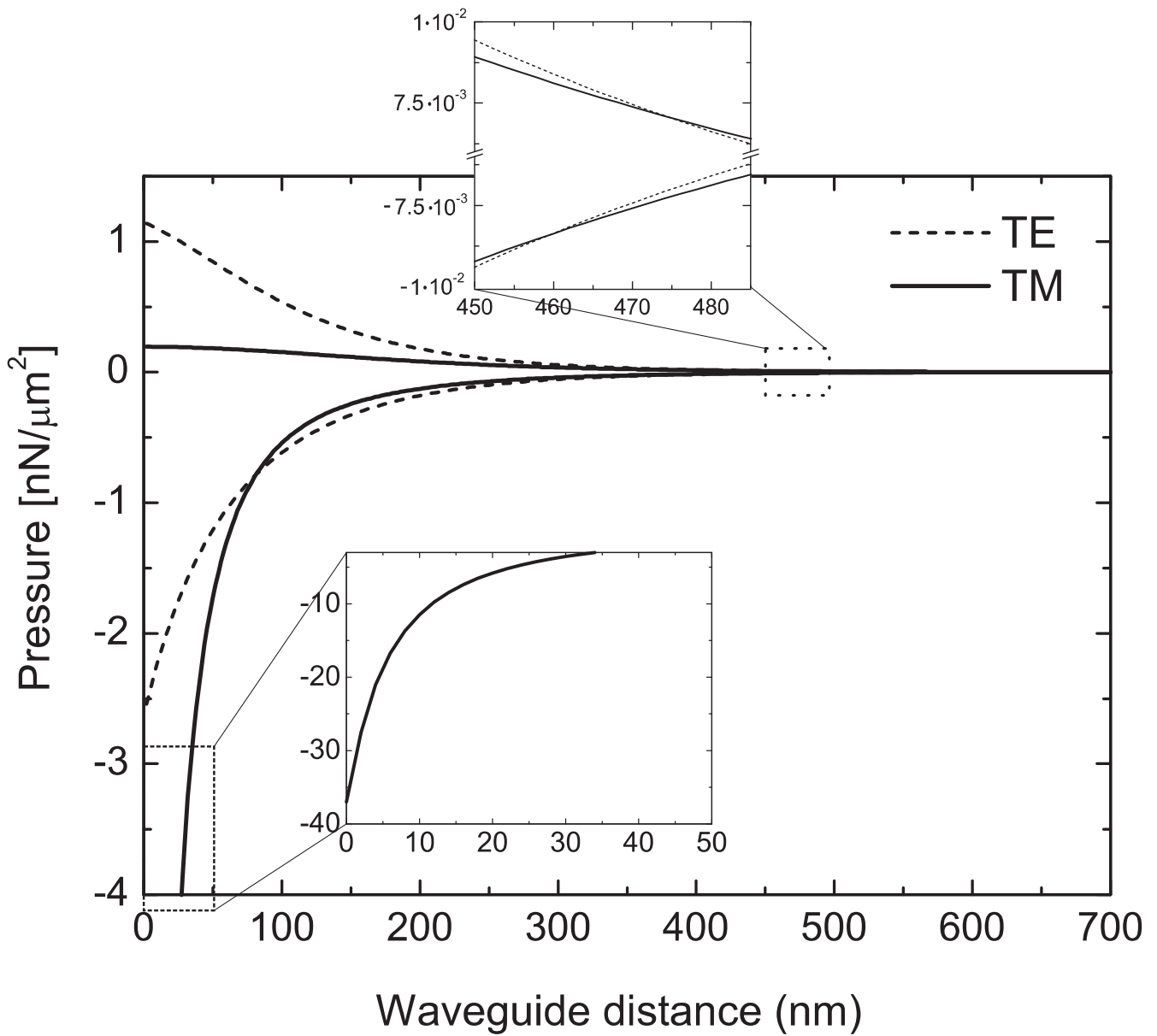}
\caption{\footnotesize Electromagnetic pressure for the fundamental ($m=0$) TE
 (continuous lines) and TM (dashed lines) modes for $s=310\,\textrm{nm}$ thick
 silicon ($n=3.5$) waveguides, a wavelength $\lambda=1.55\mu$m and a power
 density of $P=20\,\textrm{mW}/\mu\textrm{m}$. Upper (lower) curves, corresponding to a repulsive (attractive) force, refer to antisymmetric (symmetric) modes.}
\label{Force_bis}
\end{center}
\end{figure}

Thanks to the reflection symmetry of the whole set-up with respect to the 
$xz$ plane, the $y$ component to the force vanishes $F_y=0$. Also along the light propagation direction $z$ the force is zero, indeed as the geometry of the waveguide system is symmetric with respect to the 
$xy$ plane, and the dielectric medium is non-absorbing ${\rm Im}[n]=0$,
the combination of this reflection symmetry and the time-reversal is a symmetry 
of the problem. The electromagnetic force is therefore directed along the $x$
direction.
The contribution of the two planar sides parallel to the $xz$ plane
cancel each other by translational symmetry, as well as the one of the
two bases parallel to the $xy$ plane.
We are therefore left with the sides parallel to the $yz$ plane; the $x$
component of the force due to these sides only involves the $xx$
component of the Maxwell stress tensor; because of the translational
symmetry of the configuration under examination, this quantity can
only depend on the $x$ coordinate: 
\begin{equation}
\!\!\!\!\!\!\!\!\!\!\!\!\!\!\!\!\bar{T}_{xx}=
-\frac{\epsilon_0}{4}\,\left[|E_y|^2+|E_z|^2-|E_x|^2+c^2\mu_0^2\left(|H_y|^2+|H_z|^2-|H_x|^2\right)\right].
\label{MSTxxM}
\end{equation}
Inserting the explicit form of the fields, it is immediate to see that
$T_{xx}=0$ in the region $|x|>a+s$ external to the waveguide
system. This is due to the evanescent wave character of the field in this
region. 
The electromagnetic pressure $p$ (i.e., the force per unit length along $z$
and unit width along $y$) is therefore equal to $-\bar{T}_{xx}$
evaluated in the region between the two waveguides, $|x|<a$.
Positive (negative) signs for $p$ respectively indicate repulsive
(attractive) forces between the waveguides.
Plugging in (\ref{MSTxxM}) the explicit expression of the fields found
in the previous section (Sec. \ref{SecII}), we obtain the following
results for the symmetric $TE/TM$ modes:
\begin{eqnarray}
p=\frac{1}{4}\epsilon_0\,|A_{TE}|^2\left[\left(1-\frac{\beta^2}{k^2}\right)\frac{1-\tanh^2(\sigma
a)}{1+\frac{\sigma^2}{\kappa^2}\tanh^2(\sigma a)}\right],
&&\;\;\;\;\;\;\;\;\textrm{\sl TE}\label{llos}\\
\nonumber\\
p=\frac{1}{4}\mu_0\,|A_{TM}|^2\left[\left(1-\frac{\beta^2}{k^2}\right)\frac{1-\tanh^2(\sigma
a)}{1+n^4\frac{\sigma^2}{\kappa^2}\tanh^2(\sigma
a)}\right].&&\;\;\;\;\;\;\;\;\textrm{\sl TM} \label{lossd}
\end{eqnarray}
The expression for the $TE/TM$ antisymmetric modes are again found by replacing $\tanh(x)\mapsto \tanh(x)^{-1}$ in (\ref{llos})
and (\ref{lossd}). 
In the following, we will work at constant laser frequency $\omega$ and power
density $P$, so that the amplitude coefficients $A_{TE}$ and $A_{TM}$
have to be obtained by  inverting (\ref{P_TE}) and (\ref{P_TM}).

Note that since the Maxwell stress tensor is bilinear in the local fields, the
effect described here does not rely on the coherence of the light
beam, thus the results of the present paper directly extend to the
case of an incoherent, thermal source. Indeed the total pressure induced by a source with spectral distribution $f(\omega)$ is simply $\int \textrm{d}\omega\; f(\omega) p(\omega)$, where $p$ is given by Eqs. (\ref{llos}) and (\ref{lossd}).

Regarding the monochromatic source, considered hereafter, one has to distinguish two cases, depending on whether the laser frequency
is far from or close to the cut-off frequency of the considered mode.
Results for the first case are shown in Figure~\ref{Force_bis} for the
$m=0$ mode in $310$~nm thick silicon ($n=3.5$) waveguides: the
pressure is plotted as a function of the separation $2a$ between 
the waveguides.
Fixed values are taken for the power density
$P=20\,\textrm{mW}/\mu\textrm{m}$ 
and the wavelength
$\lambda=1.55\,\mu{\rm m}$ of the wave.
The main feature is that the pressure is always attractive for both
$TE$ (continuous lines) and $TM$ (dashed lines) symmetric modes, while it is
repulsive for the spatially antisymmetric modes.
Formally, this is an immediate consequence of (\ref{llos}) and
(\ref{lossd}), once one takes into account the fact that for guided
modes $\beta>k$.
In magnitude, the force is a monotonically decreasing function of the
separation $a$.
Physically, this behavior has an immediate explanation in terms of
the analogy with two coupled well models: as shown in
Fig.\ref{Band_diagram1}, the frequency of the antisymmetric
(symmetric) mode for a given $\beta$  monotonically grows (decreases) 
as the waveguides are brought closer.

\begin{figure}[ptb]
\begin{center}
\includegraphics[width=0.7\textwidth]{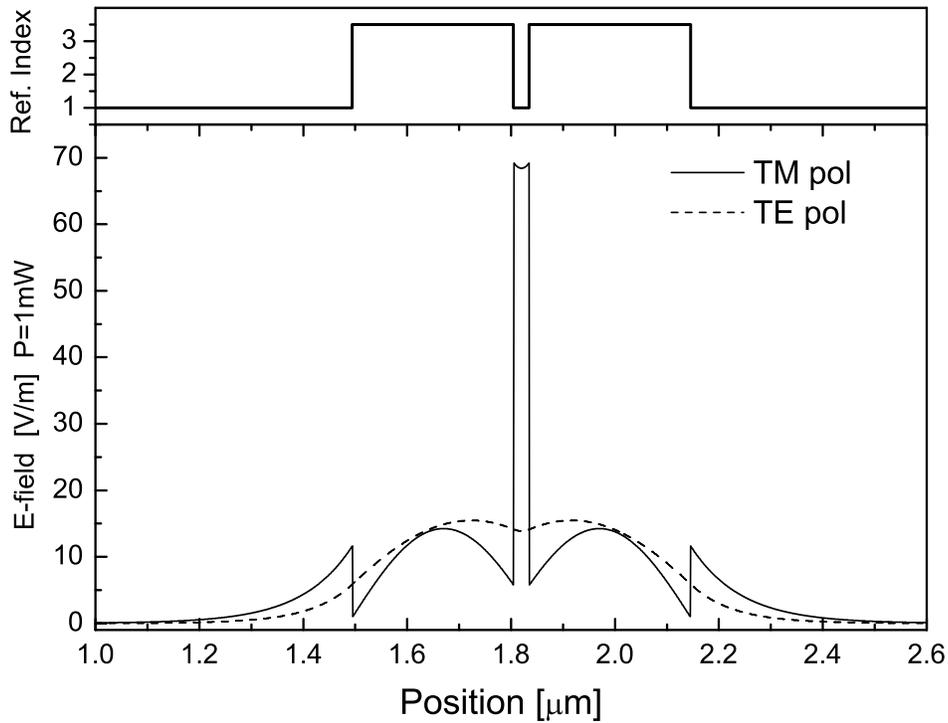}
\caption{\footnotesize Profile of the $E_x$ component of the electric field for the symmetric
 $TE$ and $TM$ $m=0$ modes. }
\label{mode}
\end{center}
\end{figure}

At large distances, the decay of the pressure as a function of
distance is exponential and for a given $TE/TM$ polarization, 
the symmetric/anti-symmetric modes only differ by the sign of the pressure.
Since the $TE$ mode is more confined in the slabs than the $TM$ one, 
it has a shorter characteristic length of the exponential.
At short distances, it is interesting to note that the $TM$ symmetric
 mode produces an significantly enhanced
pressure as compared to the corresponding $TE$ one.
A physical explanation of this behavior is readily obtained by comparing the $E_x(x)$ 
electric field profiles of the $m=0$ symmetric $TE$ and $TM$ modes, as shown in Fig.\ref{mode}: 
while the $TE$ mode profile has a smooth spatial dependence, the $TM$ one is strongly concentrated in
the region between the two slabs.
This feature is typical of $TM$ modes~\cite{Lipson}, and originates from the
continuity of the normal component $D_z$ of the electric displacement vector at the slab interface, 
which introduces a $n^2$ factor between the values of $E_z$
at the two sides of the interface. 

\begin{figure}[ptb]
\begin{center}
\includegraphics[width=1\textwidth]{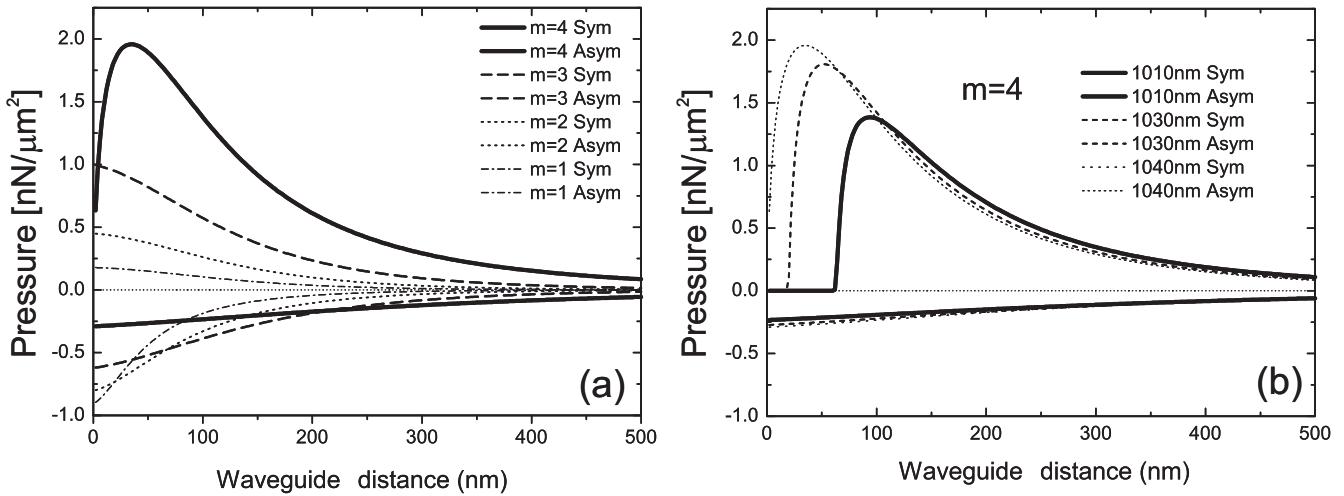}
\caption{\footnotesize Plot of the pressure as a function of the separation $a$ for 
(a) modes with a different order $m$;
(b) for the same $m=4$ and slightly
 different waveguide thickness. Upper (lower) curves, corresponding to a repulsive (attractive) force, refer to antisymmetric (symmetric) modes.}
\label{Figure5_ab_bis}
\end{center}
\end{figure}

If the laser frequency is not far from the cut-off of the mode, the
dependence of the pressure $p$ on the separation $a$ is somehow richer.
In Fig.\ref{Figure5_ab_bis}a we consider the case of a thicker waveguide 
$s\approx 1~\mu m$.
For the wavelength $\lambda=1.55\,\mu{\rm m}$ under consideration, 
all the modes up to $m=3$ are well confined, while we are 
just above the isolated ($a/s=\infty$) waveguide cut-off for the $m=4$ mode.
Since the cut-off frequency of anti-symmetric modes increases for
decreasing $a$, it exists a cut-off separation $a_{co}$
below which light of the given wavelength ceases being guided
in the $m=4$ mode. 
When $a\rightarrow a_{co}$ from above, the spatial size $\sigma^{-1}$ of
the mode diverges, and the field amplitude between the guides tends
(for a given power density $P$) to zero.
As a consequence, the pressure $p$ initially grows for decreasing $a$, 
attains a maximum value at some separation value, and then goes back to zero 
as the cut-off separation $a_{co}$ is approached~\cite{note}.
Clearly, the cut-off separation is larger for thinner waveguides
(Fig.\ref{Figure5_ab_bis}b).
This behavior can also be explained in terms of the two coupled
well model: when the eigenstates of the independent wells are
close to the continuum threshold, there exists a value of the distance (i.e., of the coupling strength), 
at which the antisymmetric state ceases to be bound and enters the continuum.
On the other hand, for the tightly confined $m\leq 3$ modes, the physics
is qualitatively the same as in Fig.\ref{Force_bis}.

\section{Radiation induced deflection: the case of the air-bridge waveguide}
\label{airbridge}
Although the value of the radiation pressure obtained in the previous section might seem at a first glance rather small, it can have a observable effect in nanometric devices. 
In order to provide a quantitative estimate of such effects we consider the mechanical deflection induced by the optical force on an air-bridge double slab waveguide system made of two thin silicon slabs of length $L$ whose opposite edges are clamped to the substrate. 
Such a device can be realized, e.g., {\sl via} chemical etching technique in \cite{sacrificiallayer}.

\begin{figure}[ptb]
\begin{center}
\includegraphics[width=0.8\textwidth]{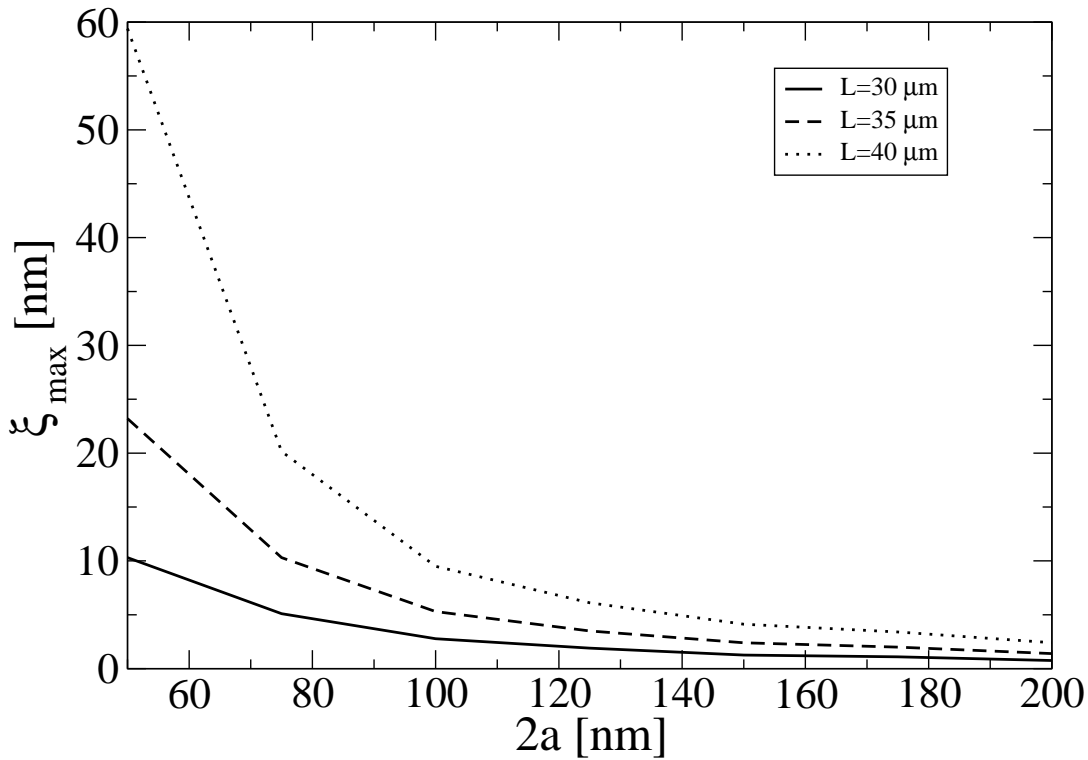}
\caption{\footnotesize Maximal value $\xi_{\textrm{max}}$ of the deflection $\xi$ produced by the TM symmetric mode (with power $P=20\,\textrm{mW}/\mu\textrm{m}$ and wavelength $\lambda=1.55\mu$m) of the radiation induced force on an silicon double air-bridge waveguide of thickness $s=310$nm, as a function of the initial separation $2a$. Three different values of the  slabs length are considered: $L=30 \mu$m (solid), $L=35 \mu$m (dashed) and $L=40 \mu$m (dotted).}
\label{Figure_defl}
\end{center}
\end{figure}

The deflection $\xi$ induced  by the radiation pressure on the device 
can be evaluated using the Euler-Bernoulli beam equation \cite {LandauElastic}
\begin{equation}
E I\frac{\partial ^4 \xi}{\partial x^4}=h p(\xi),
\label{dddd}
\end{equation}
where, $\xi$ is the separation between the slabs, $E$ is the Young modulus, and  $I=(1/12) h s^3$ is the area moment of inertia of the slab's cross section, whose width and thickness are $h$ and $s$. The maximum deflection $\xi_{max}$ induced on each silicon  ($E=169$GPa) slab by a power $P=20\,\textrm{mW}/\mu\textrm{m}$ at a wavelength $\lambda=1.55\mu$m propagating in the TM symmetric mode is shown in Fig.(\ref{Figure_defl}) as a function of the initial separation $2a$ for three different values of the slab length $L=30,\,35,\,40\, \mu$m, and for slab's thickness $s=310$nm. For this parameter choice, a linearized treatment of Eq.(\ref{dddd}) around the initial slab distance already provides accurate results.

From figure (\ref{Figure_defl}) one can see that the maximal value $\xi_{\textrm{max}}$ of the deflection $\xi$ induced by the optical force can reach the values of tens nanometers for reasonable structural parameters:  slabs lengths $30 \mu$m$<L<40 \mu$m (the deflection is strongly sensitive to $L$), and  waveguides separations $50$nm$<2 a< 200$nm. Such deflections are of the same order of previous investigations made for different configurations \cite{povinelli} and are accessible to standard Atomic Force Microscope techniques~\cite{AFM}. 
As a final remark, as the pressure is roughly speaking inversely proportional $p\propto 1/v_g$ to the group velocity $v_g$ of the mode, one expects that the radiation pressure, and hence the mechanical deflection, can be enhanced if the $v_g$ is slowed down by a longitudinal modulation of the structure as in coupled resonator optical waveguides \cite{CROW}.
%
\section{Large and small distance behavior}
\label{sec:limits}
%
In this section we derive the asymptotic behavior of the pressure for small and large slab separation, which provides a deeper insight on the findings of the previous section. 

\subsection{Large distance behavior}
\label{sec:large}

For  large separation $a/s\rightarrow \infty$, the force
has the typical exponential decay of two-well systems in the tight-binding limit.
As long as the modes are confined, the general qualitative trend is that the larger the characteristic length $(\sigma^{TE/TM}_\infty)^{-1}$, the larger the
field in between the slabs and consequently the stronger the force.
More specifically: for a given order $m$ the pressure is (slightly) stronger for the $TM$ mode than for the corresponding $TE$ one.
(ii) The pressure is stronger for higher $m$ modes.

Quantitative expressions for the case of symmetric (anti-symmetric)
modes can be obtained by expanding Eqs. (\ref{llos}) and (\ref{lossd})
for large distances: 
\begin{eqnarray}
\!\!\!\!\!\!\!\!\!\!\!\!\!\!\!\!\!\!\!\!\!\!\!\!\!\!p_{tb}^{TE,\pm}=\pm P\left[\frac{k
    \left(1-\frac{\beta^{TE\;2}_\infty}{k^2}\right)}{c\beta^{TE}_\infty 
s(1+\frac{2}{\sigma^{TE}_\infty s})(1+\frac{\sigma^{TE\;2}_\infty}{\kappa^{TE\;2}_\infty})}\right]\,
e^{-2a\sigma^{TE}_\infty}
\label{largedist1}\\ 
\!\!\!\!\!\!\!\!\!\!\!\!\!\!\!\!\!\!\!\!\!\!\!\!\!\!p_{tb}^{TM,\pm}=\pm P\left[\frac{k\,n^2\left(1-\frac{\beta^{TM\;2}_\infty}{k^2}\right)}{c\beta^{TM}_\infty  
s\left(1+\left(\frac{n^2\sigma^{TM}_\infty}{\kappa^{TM}_\infty}\right)^2+\frac{2n^2}{\sigma^{TM}_\infty s}\left(1+\frac{\sigma^{TM\;2}_\infty}{\kappa^{TM\;2}_\infty} 
\right)\right)}\right]e^{-2a\sigma^{TM}_\infty}.\label{largedist}
\end{eqnarray}
where $\kappa^{TE/TM}_\infty$, $\sigma^{TE/TM}_\infty$, $\beta^{TE/TM}_\infty$ are evaluated at infinite separation $a/s=\infty$. As before, the $+$($-$) sign refers to the symmetric (antisymmetric) mode.

\subsection{Small distance behavior}
\label{sec:short}

In order to get some analytical insight in the small distance regime, it
is useful to expand all the waveguide parameters in powers of the slab separation $a$,
while keeping $P$ and $\omega$ constant.
Let us consider the specific case of $\beta(a) = \beta_0+\beta_1\,a+O(a^2)$.
The zeroth order $\beta_0=\beta(a=0)$ is the wave number of
the propagation along a single waveguide of double thickness $2s$, and
has to be calculated from the dispersion laws 
once the hyperbolic tangent is replaced by its limiting value $1$ and
the doubled thickness is taken into account as $s\rightarrow 2s$.
The first order term $\beta_1$ is given, e.g., for $m=0$ symmetric modes by:
\begin{eqnarray}
\!\!\!\!\!\!\!\!\!\!\!\!\beta_1(a)=-\frac{(\sigma^{TE,+}_0)^3}{\beta^{TE,+}_0\,(1+s\sigma^{TE,+}_0 )},&&\;\;\;\;\;\;\;\;\;\;\;\;\;\;\textrm{\sl TE}\label{beta1TE}\\
\!\!\!\!\!\!\!\!\!\!\!\!\beta_1(a)=-\frac{(\sigma_0^{TM,+})^3}{\beta_0^{TM,+}\left[\frac{(\sigma_0^{TM,+})^2+(\kappa_0^{TM,+})^2}{(\sigma^{TM,+}_0 
n^2)^2+(\kappa_0^{TM,+})^2}+\frac{\sigma_0^{TM,+}\,s}{n^2}\right]}.&&\;\;\;\;\;\;\;\;\;\;\;\;\;\;\textrm{\sl TM}
\label{beta1TM} 
\end{eqnarray}
Note that $\beta_1$ is negative for all modes, indicating that the force is attractive.
Along the same lines, analytical expansions can be obtained for the pressure $p$.
The $a=0$ value $p_0$ for the symmetric $TE/TM$ modes has the form
\begin{eqnarray}
p_0^{TE,+}=P\;\frac{(\sigma_0^{TE,+})^3}{2\omega\beta_0^{TE,+}(1+\sigma_0^{TE,+}s)}, \label{min}\\
p_0^{TM,+}=P\;\frac{(\sigma_0^{TM,+})^3} {2\omega\beta_0^{TM,+}
\left[\frac{(\kappa_0^{TM,+})^2+(\sigma_0^{TM,+})^2}{(\kappa_0^{TM,+})^2+n^4(\sigma_0^{TM,+})^2}+\frac{\sigma_0^{TM,+}
s}{n^2}\right]}.
\label{force0sym}
\end{eqnarray}
Analogous algebra leads to the corresponding expressions for the anti-symmetric $TE/TM$ modes, which have the form:
\begin{eqnarray}
p_0^{TE,-}=-P\;\frac{\sigma^{TE,-}_0(\kappa_0^{TE,-})^2}{2\omega\beta^{TE,-}_0(1+\sigma^{TE,-}_0s)},\label{fff}\\
p_0^{TM,-}=-P\;\frac{\sigma^{TM,-}_0\,(\kappa_0^{TM,-})^2} {2n^4\,\omega\beta^{TM,-}_0
\left[\frac{(\kappa^{TM,-}_0)^2+(\sigma^{TM,-}_0)^2}{(\kappa_0^{TM,-})^2+n^4(\sigma_0^{TM,-})^2}+\frac{\sigma^{TM,-}_0
s}{n^2}\right]}.
\label{force0anti}
\end{eqnarray}
Starting from these formulas, a physical explanation can be provided for the remarkable facts observed in Fig.\ref{Band_diagram1} 
for the $m=0$ modes, namely the suppressed value of the force for the anti-symmetric $TM$ mode and the enhanced value 
for the symmetric $TM$ mode with respect to the $TE$ modes.
As long as we are considering well confined modes, the $\sigma_0$'s of all modes have almost comparable values, somewhat larger than the $\kappa_0$'s. This explains the general fact that the force is about a factor $2$ weaker for the $TE$ antisymmetric mode than for the corresponding symmetric one.
The behavior for the $TM$ modes can be explained starting from the value $n=3.5$ of the refractive index, which is significantly larger than $1$:  thanks to the $n^4$ in the denominator, the force  $p_0^{TM,-}$ for the antisymmetric $TM$ mode is dramatically suppressed of a factor  $\approx 150$ (for the chosen value $n=3.5$ of the refractive index) with respect to the one $p_0^{TM,+}$ for the symmetric $TM$ mode.
For similar reasons, the force $p_0^{TM,+}$ is enhanced of a factor $\approx 13$ over $p_0^{TE,+}$ because of the $n$'s in the denominator (see Fig. \ref{Force_bis}).

\section{Conclusions}
\label{conclu}

In conclusion, we have performed an analytical study of the optical
forces appearing between a pair of parallel slab waveguides when light
is propagating through them.
Depending on the spatial symmetry of the mode wavefunction, the sign
of the force can be either attractive or repulsive.
The dependence of the force as a function of the separation between
the slabs has been characterized for the different polarization
states, and analytical expressions have been obtained for both the
large and the small distance limits.
A strong enhancement of the force has been identified for the
symmetric $TM$ mode, as well as a suppression for the antisymmetric
$TM$ one. Simple physical explanations have been provided for these
features. A quantitative study for typical air-bridge configurations confirms 
that the mechanical deflection of the structure induced by the optical force 
can be measured by standard Atomic Force Microscopy techniques.

\section{Acknowledgment}
We thank L. P. Pitaevskii and D.S. Wiersma for useful discussions and we acknowledge support by the Ministero dell'Istruzione, dell'Universit\`a e della Ricerca (MIUR).



\begin{thebibliography}{10}
\bibitem{note}
Note that this non-monotonic behavior has a completely different
origin from the one reported in~\cite{povinelli}.
There, a pair of linear waveguides were considered instead of slab
ones, and the change of sign of the pressure for the antisymmetric
TE mode was due to edge effects because of the confinement in the
$y$ direction.
\end{thebibliography}
\end{document}